\documentclass[atom,preprint,review,submit,moreauthors,pdftex,10pt,a4paper]{Definitions/mdpi} 
\usepackage{amsmath,amssymb}
\usepackage{hyperref}
\DeclareFontFamily{U}{rcjhbltx}{}
\DeclareFontShape{U}{rcjhbltx}{m}{n}{<->rcjhbltx}{}
\DeclareSymbolFont{hebrewletters}{U}{rcjhbltx}{m}{n}
\let\aleph\relax\let\beth\relax
\let\gimel\relax\let\daleth\relax
\preto{\abstractkeywords}{\nolinenumbers}
\DeclareMathSymbol{\aleph}{\mathord}{hebrewletters}{39}
\DeclareMathSymbol{\beth}{\mathord}{hebrewletters}{98}
\DeclareMathSymbol{\gimel}{\mathord}{hebrewletters}{103}
\DeclareMathSymbol{\daleth}{\mathord}{hebrewletters}{100}
\DeclareMathSymbol{\lamed}{\mathord}{hebrewletters}{108}
\DeclareMathSymbol{\mem}{\mathord}{hebrewletters}{109}
\DeclareMathSymbol{\ayin}{\mathord}{hebrewletters}{96}
\DeclareMathSymbol{\tsadi}{\mathord}{hebrewletters}{118}
\DeclareMathSymbol{\qof}{\mathord}{hebrewletters}{113}
\DeclareMathSymbol{\shin}{\mathord}{hebrewletters}{152}
\def\aap{A\&A\/}
\def\apjs{ApJS}
\def\apj{ApJ}
\def\aj{AJ}
\def\nat{Nat}
\def\apjl{ApJL}
\def\mnras{MNRAS}
\def\pasj{PASJ}
\def\nar{NAR}
\def\araa{ARA\&Ap}
\firstpage{1} 
\pubvolume{xx}
\issuenum{1}
\articlenumber{5}
\pubyear{2018}
\copyrightyear{2018}
\history{Received: date; Accepted: date; Published: date}
\Title{Quasars: from the Physics of Line Formation to Cosmology}

\Author{P. Marziani$^{1}$\orcidA{},  E. Bon$^{2}$, N. Bon$^{2}$,  A. del Olmo$^{3}$, M. L. Mart\'{\i}nez-Aldama$^{3}$,  M. D'Onofrio$^{4}$, D. Dultzin$^{5}$, C. A. Negrete$^{5}$,  G.  M. Stirpe$^{6}$}

\AuthorNames{Paola Marziani, Edi Bon and Natasa Bon}

\address{%
$^{1}$\quad National Institute for Astrophysics (INAF), Astronomical Observatory of Padova, IT 35122, Padova, Italy; paola.marziani@inaf.it\\
$^{2}$\quad Astronomical Observatory, Belgrade, Serbia\\ 
$^{3}$\quad Instituto de Astrofis\'{\i}ca de Andaluc\'{\i}a, IAA-CSIC, Glorieta  de la Astronomia s/n, E-18008 Granada, Spain\\
$^{4}$\quad Dipartimento di Fisica \& Astronomia ``Galileo Galilei'', Universit\`a di Padova, Padova,  Italy \\
$^{5}$\quad Instituto de Astronom\'{\i}a, UNAM, Mexico D.F. 04510, M\'exico\\
$^{6}$\quad INAF, Osservatorio di astrofisica e scienza dello spazio, Bologna, Italy}
\abstract{Quasars accreting matter at very high rates (known  as extreme Population A [xA] or super-Eddington accreting massive black holes)  provide a new class of distance indicators covering cosmic epochs  from the present-day Universe up to less than 1 Gyr  from the Big Bang. The very high accretion rate  makes it possible that  massive black holes hosted in { xA} quasars radiate at a stable, extreme luminosity-to-mass ratio.  This in turns translates into  stable physical and dynamical conditions of the mildly ionized gas in the quasar low-ionization line emitting region. In this contribution, we  analyze the main optical and UV spectral properties of  extreme Population A  quasars  that make them easily identifiable in large spectroscopic surveys  at low- ($z \lesssim 1$) and intermediate-$z$ (2  $\lesssim z \lesssim$ 2.6), and  the physical conditions that are derived for the formation of their emission lines. Ultimately, the  analysis supports the possibility of identifying a virial broadening estimator from low-ionization line widths, and  the conceptual validity of the redshift-independent luminosity estimates based on virial broadening for a known luminosity-to-mass ratio. }

\begin{document}

\section{Introduction}

\subsection{Quasar spectra: emission from mildly ionized gas}

The spectra of quasars can be easily recognized  by the presence of broad and narrow optical and UV lines emitted by mildly- ionized species over a wide range of ionization potential. The  type-1  composite quasar spectrum from the SDSS \cite{vandenberketal01} reveals  Broad (FWHM $\gtrsim $ 1000 km s$^{-1}$)\ and Narrow High Ionization lines (HILs, ~ 50eV) and Low Ionization lines (LILs, < 20eV).  Broad HILs encompass  CIV$\lambda$1549, HeII$\lambda$1640 and HeII$\lambda$4686  as representative specimens. Broad LILs include  HI Balmer lines (H$\beta$, H$\alpha$), MgII$\lambda$2800, the CaII IR Triplet, and FeII features. The FeII emission deserves a particular mention, as it is extended over a broad range of wavelengths (Fig. 6 of \cite{vandenberketal01}), and is especially prominent around MgII$\lambda$2800\ and H$\beta$.  The FeII emission is one of the dominant coolants in the broad line region (BLR)  and therefore a main factor in its energetic balance ({ the FeII emission extends from the UV to the FIR \cite{marzianietal06}, and can reach the luminosity of Ly$\alpha$, \cite{netzer90,marinelloetal16}}). So it may not appear surprising that an estimator of its strength plays an important role in the systematic organization of quasar properties (\S \ref{e1xa}).   

This paper reviews results obtained in the course of  two decades (\S \ref{e1xa} and \S \ref{phys}), attempting to explain how the spectral properties  of a class of type-1 quasars and their physical interpretation can lead to the definition of ``Eddington standard candles'' (ESC, \S \ref{esc}). { In the following, we will restrict the presentation to type-1 quasars which are considered mainly ``unobscured'' sources with an unimpeded view of the BLR, and exclude type-2 AGN of quasars in which the broad lines are not detected in natural light (see \cite{antonucci93} for an exhaustive review).}  We  describe the physical basis of the method in \S \ref{phys} and \S\ \ref{virial}. We then introduce  ESC selection criteria (\S \ref{selec}) and  preliminary cosmology results (\S \ref{cosmo}) . 



\subsection{Quasars for cosmology: an open issue}\label{qc}

The distribution of quasars in space and the intervening absorptions along the line of sight (i.e., the so called Ly$\alpha$\ forest) have been since long considered as tracer of matter in the distant Universe (see \cite{donofrioburigana09} and references therein).  However, a relevant question may be why  intrinsic properties of quasars have never been successfully used as cosmological probes. On the one hand, (1) quasars are easily recognizable and  plentiful ($\gtrsim$500,000 in the  data release 14 of the SDSS, \cite{parisetal18}). (2) They are very luminous,  can reach bolometric luminosity $L \gtrsim 10^{48}$\ erg s$^{-1}$; (3) they are observed in an extremely broad range of redshift $0 \lesssim z \lesssim 7$, and (4) they are  stable compared to transients that are employed as distance indicators in cosmology, such as type Ia supernov\ae\ { (Sect. \ref{def}, \cite{marzianietal18} for a review)}.   On the other hand, (1) quasars are anisotropic sources { even if the degree of anisotropy is expected to be associated with the viewing angle of the accretion disk in radio-quiet quasars \cite{negreteetal18}, and not large compared to radio-loud quasars whose optical continuum is in part beamed (see, for example \cite{liuetal06})};  (2) quasars have an open-ended luminosity function ({ i.e., without a clearly defined minimum, as the quasar highest spatial density occurs at the lowest luminosity}); in other words, they are the ``opposite'' of a cosmological standard candle. In addition, (3) the long-term variability of radio-quiet quasars is poorly understood (see e.g. \cite{bonetal16,bonetal17} and references therein) (4) and the internal structure of the active nucleus ($\lesssim$ 1000 $r_\mathrm{g}$) is still a matter of debate (see e.g. a summary of open issues \cite{netzer18} in \citep{marzianietal18x}. 
Correlations with luminosity have been proved to be rather weak (see \cite{sulenticetal00a}, for a synopsis up to mid-1999).  Selection effect may even cancel out the  ``Baldwin effect'' \cite{baldwinetal78}, a significant but weak anti-correlation between rest-frame equivalent width and continuum luminosity of  CIV$\lambda$1549 that has been the most widely discussed luminosity correlation in the past decades. 

\begin{figure}[H]
\begin{minipage}[t]{0.4\linewidth}
\centering
\includegraphics[width=7 cm]{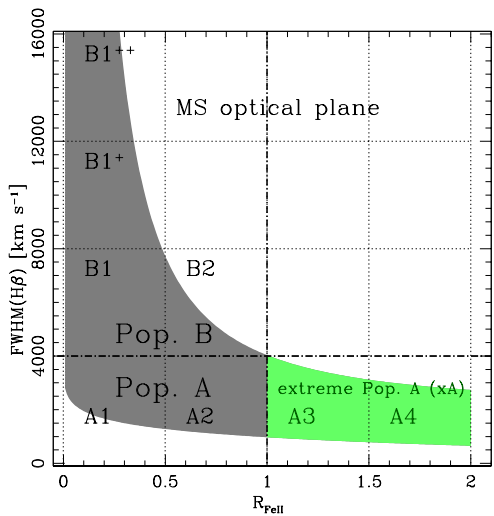}
\end{minipage}\hspace{1cm}
\begin{minipage}[t]{0.5\linewidth}
\centering
\vspace{-7cm}
\caption{The plane FWHM(H$\beta$) vs $R_\mathrm{FeII}$. The MS is sketched as the grey strip, with the section occupied by xA sources colored pale green. The thick dot-dashed line separates Pop. A and B at 4000 km s$^{-1}$, while the vertical one at $R_\mathrm{FeII}$ = 1 traces the $R_\mathrm{FeII}$\ lower value for xA identification. The spectral types with significant occupation at low-$z$ are labeled.   \label{fig:e1fspig}}
\end{minipage}
\end{figure}   


\section{Definition of a class of type-1 quasars with properties of  Eddington standard candles}\label{e1xa}
\label{def}
Nonetheless, new developments in the past decades have paved the road to the possibility of exploiting quasars as cosmological distance indicators in a novel way that would make them literal  ``Eddington standard candles'' (ESC) (\cite{teerikorpi11,wangetal13,wangetal14,marzianisulentic14}; see also \cite{czernyetal18} for a comprehensive review of secondary distance indicators including several techniques based on quasars).  This possibility is based in the development of the concept of  a quasar main sequence (MS), intended to provide a sort of H-R diagram for quasars \cite{sulenticetal08}.   The quasar MS can be traced in the plane  defined by the prominence of optical FeII emission,   $R_\mathrm{FeII}$ = I(FeII$\lambda$ 4570)/I(H$\beta$) (see \cite{borosongreen92,sulenticetal00a,sulenticetal02,zamfiretal10,shenho14}). { Fig. \ref{fig:e1fspig} provides a sketch of the MS in the optical plane FWHM(H$\beta$) vs. $R_\mathrm{FeII}$.} It is possible to isolate spectral types in the optical plane of the MS as a function of  $R_\mathrm{FeII}$ and FWHM H$\beta$ and,  at a coarser level, two  populations: Population A (FWHM H$\beta <$4000 km/s) and Population B of broader sources.  Pop. A is rather heterogeneous, and encompasses a range of $R_\mathrm{FeII}$\ from almost 0 to the highest values observed ($R_\mathrm{FeII} \gtrsim 2$\ are very rare, $\lesssim$ 1\%\ in optically-selected samples, \cite{zamfiretal10}). 
{ Along the quasar main sequence, the extreme Population A (xA) sources {  satisfying the condition $R_\mathrm{FeII} > 1$\   (about $10 $ \%\ of all quasars in optically-selected sample, green area in Fig. \ref{fig:e1fspig})} show remarkably low optical variability, so low that it is even  difficult to estimate the BLR radius via reverberation mapping \cite{duetal18}. This is at variance with Pop. B sources that show 
more pronounced variability \cite{dultzin-hacyanetal92,giveonetal99}, the most extreme cases being observed among blazars which are low-accretors,  at the opposite end in the quasar MS.}   Of the many multi-frequency trends along the main sequence (from the  sources whose spectra show the broadest LILs [extreme Pop. B], and the weakest FeII emission, to sources with the  narrowest LIL profiles  and strongest FeII emission [extreme Pop. A]), we recall a systematic decrease of the CIV  equivalent width, an increase in metallicity, and amplitude of HIL blueshifts (a more exhaustive list is provided by Table 1 of \cite{sulenticetal11}). Eddington ratio is believed to increase along with $R_\mathrm{FeII}$\ \citep{borosongreen92,kuraszkiewiczetal04,shenho14,sunshen15}. The FWHM H$\beta$ is strongly affected by the viewing angle (i.e., the angle between the line of sight and the accretion disk axis), so that  at least most narrow-line Seyfert 1s (NLSy1s) can be interpreted as Pop. A sources seen with the accretion disk oriented face-on or almost so \cite{dultzinetal11}.  At low-$z$ ($\lesssim$0.7), Pop. A implies low black hole mass $M_\mathrm{BH}$, and high Eddington ratio; on the converse, Pop. B is associated with high $M_\mathrm{BH}$\ and low $L/L_\mathrm{Edd}$.   This trend follows from  the ``downsizing'' of nuclear activity at low-$z$ that helps give an elbow shape to the MS \cite{fraix-burnetetal17}: at low-$z$, very massive quasars ($M_\mathrm{BH}\gtrsim 10^{9}$\ M$_{\odot}$) do not radiate close to their Eddington limit but are, on the converse, low-radiators ($L/L_\mathrm{Edd} \lesssim 0.1$).  

The inter-comparison between CIV$\lambda$1549 and H$\beta$ supports low-ionization lines virial broadening (in a system of dense clouds or in the accretion disk) + high-ionization lines (HILs) radial or vertical outflows, at least in Pop. A sources \cite{leighly04,sulenticetal17}. There is now a wide consensus on an accretion disk + wind system model \cite{elvis00}, and therefore on the existence of  a ``virialized'' low-ionization subregion + higher ionization, outflowing subregion up to the highest quasar luminosities \cite{bisognietal17,sulenticetal17,vietrietal18}. 


The most extreme examples at high accretion rate are a population of sources with distinguishing properties. They have been called extreme Pop. A or extreme quasars (xA), and are also known as super-Eddington accreting massive black holes (SEAMBHs) \cite{negreteetal18,martinez-aldamaetal18,wangetal13,duetal16}.  Observationally, xA quasars satisfy $R_\mathrm{FeII} \ge 1$\ and still show LIL H$\beta$\ profile consistent with emission from a virialized system.  xA quasars may be well represent an early  stage in the evolution of quasars and galaxies. In the hierarchical growth scenario for the evolution of galaxies,   merging and strong interaction lead to accumulation of gas in the galaxy central regions, inducing   enhanced star formation. { Strong winds from massive stars and eventual} Supernova explosions { may ultimately provide enriched} accretion fuel for the massive black hole at the galaxy \cite{hellershlosman94,collinzahn99,williamsetal99}. The active nucleus radiation force and the  mechanical thrust of the accretion disk wind can then sweep the dust surrounding the black hole, at least within a cone coaxial with the accretion disk axis (see Fig. 7 of  \cite{donofriomarziani18}). { The fraction of mass that is accreted by the black hole and the fraction that is instead ejected in the wind are highly uncertain; the outflow kinetic power can become comparable to the radiative output \cite{kingpounds15,marzianietal17c}, especially in sources accreting at very high rate \cite{nardinietal15}; interestingly, this seems to be true also for stellar-mass black holes \cite{tetarenkoetal18}.} Feedback effects on the host galaxies are maximized by the high kinetic power of the wind, presumably made of gas much enriched in metals \cite{baskinlaor12}.

\begin{figure}
\centering
\includegraphics[scale=0.55]{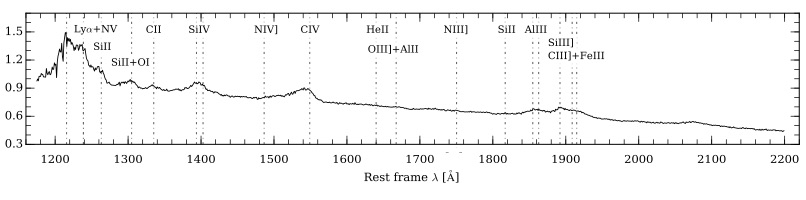}
\vspace{-0.75cm}
\caption{Composite UV spectrum of high-$z$\ xA sources. Abscissa is rest-frame wavelength, ordinate is normalized flux.   \label{fig:uvsp} }
\end{figure}

\section{Diagnostics of mildly-ionized gases }
\label{phys}
 
Diagnostics from the rest-frame UV spectrum takes advantage of the observations of strong resonance lines that are collisionally excited \cite{negreteetal12,negreteetal13}. The point is that the rest-frame UV spectrum offers a rich diagnostics that constrains at least gas density $n_\mathrm{H}$, ionization parameter $U$, chemical abundance $Z$.  For instance, Si II$\lambda$ 1814/Si III]$\lambda$  1892 sensitive to ionization  CIV$\lambda$1549/Ly$\alpha$, CIV$\lambda$1549/(Si IV+OIV])$\lambda$1400, CIV$\lambda$1549/HeII$\lambda$1640,  NV$\lambda$1240/HeII$\lambda$1640 are sensitive to metallicity; Al III$\lambda$1860/Si III]$\lambda$1892, Si III]$\lambda$1892/CIII]$\lambda$1909 are sensitive to density, since inter-combination lines have a well defined critical density \citep{negreteetal12}. 

The photoionization code {\em Cloudy} models the ionization, chemical, and thermal state of gas exposed to a radiation field, and predicts its emission  spectra and physical parameters \cite{ferlandetal13,ferlandetal17}. In {\em Cloudy}, collisional excitation and radiative processes typical of mildly ionized gases are included. {\em Cloudy} simulation require inputs in terms of  $n_\mathrm{H}$,  $U$, $Z$, quasar spectral energy distribution (SED), column density $N_\mathrm{c}$.   The ionization parameter

\begin{equation}
U = \frac{\int_{\nu_{0}}^{\infty} \frac{L_{\nu}}{h\nu}  }{4 \pi r_\mathrm{BLR}^{2} c n_\mathrm{H}} = \frac{Q(H)}{4 \pi r_\mathrm{BLR}^{2} c n_\mathrm{H}}, \label{eq:ip}
\end{equation}
where $Q(H)$ is the number of ionizing photons, provides the ratio between photon and hydrogen number density. More importantly, the inversion of equation provides a measure of the emitting region radius $r_\mathrm{BLR}$ once  the ionizing photon flux i.e., the product $U n_{H}$\ is known. As we will see, the photon flux can be estimated with good precision from diagnostic line intensity ratios. 

\begin{figure}
\centering
\includegraphics[scale=0.35]{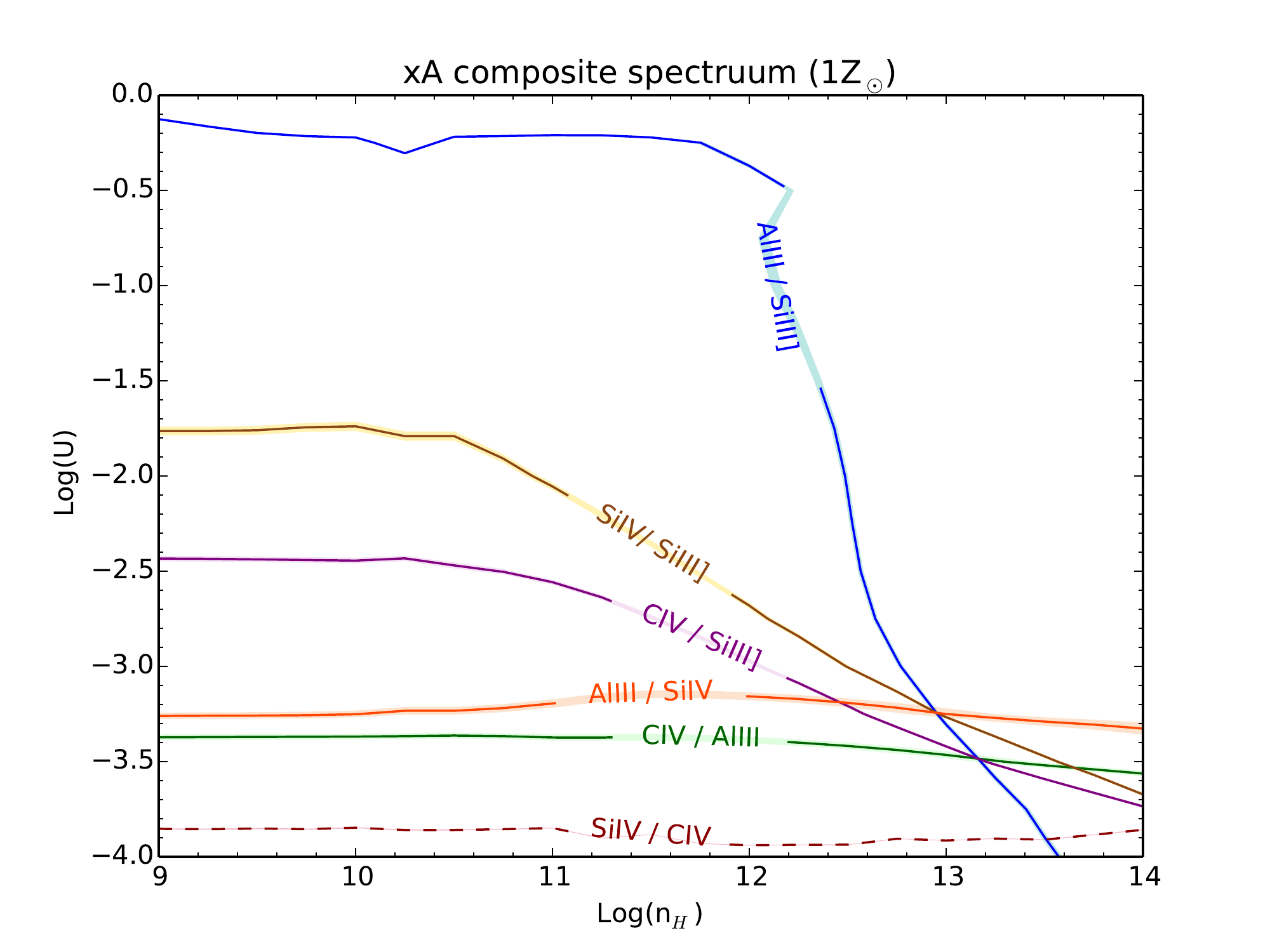}
\includegraphics[scale=0.35]{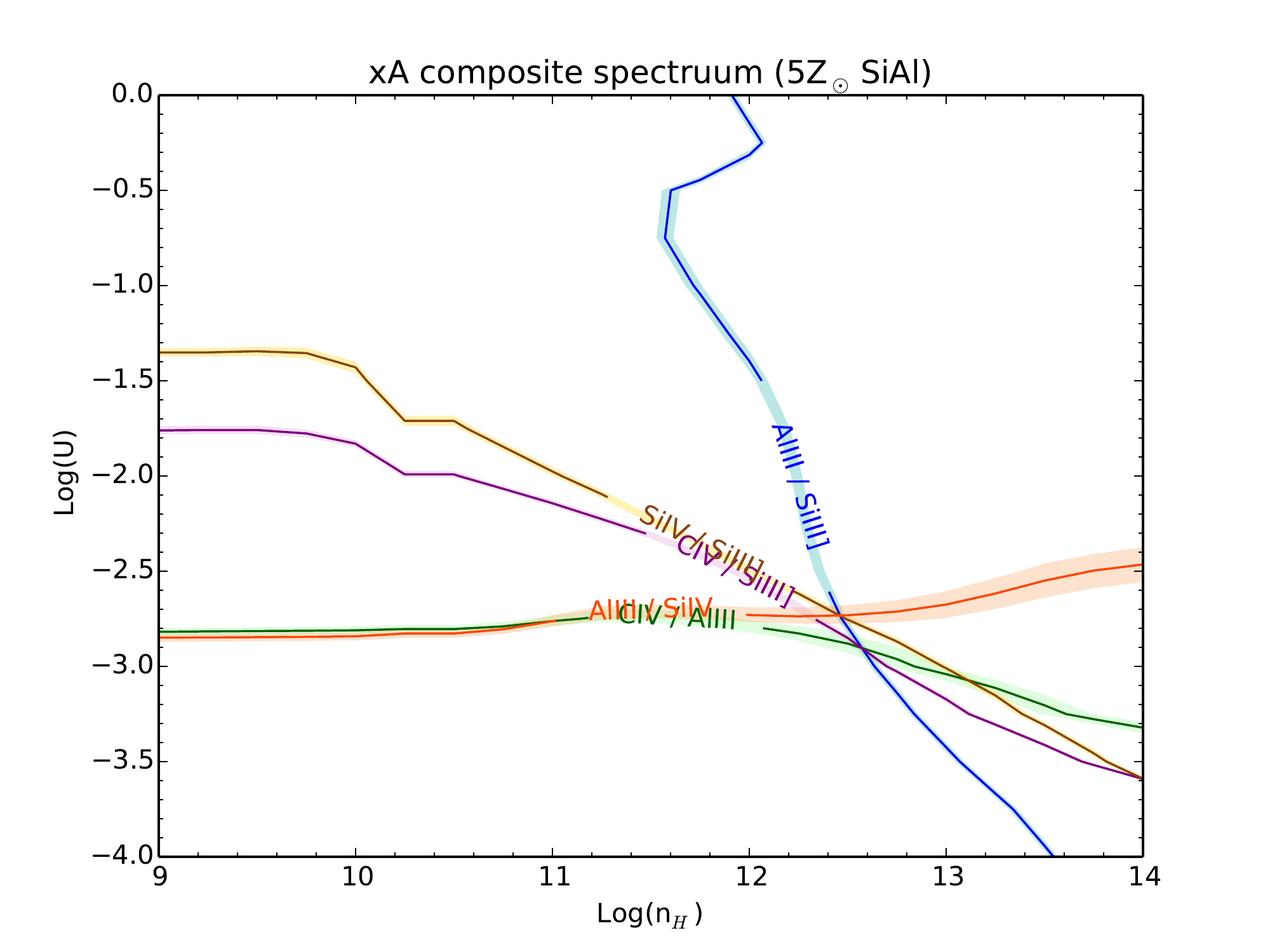}\\
\caption{Intensity ratios in the plane  ionization parameter vs. density, for the intensity ratios measured on the composite xA quasar spectrum shown in Fig. 2 of  \cite{martinez-aldamaetal18}. Left panel: solar chemical composition; right: $5\times$ solar chemical composition with selective enrichment in Al and Si, following \cite{negreteetal12}. In this latter case the SiIV$\lambda$1402/CIV$\lambda$1549 is degenerate.   \label{fig:diag} }
\end{figure}

Maps built on an array of 551 Cloudy 08.00 -- 13.00  photoionization models for a given metallicity $Z$\ and $N_\mathrm{c}$, constant $n$\ and $U$\ evaluated at steps of 0.25 dex covering the ranges $7 \le \log n_\mathrm{H} \le 14$ [cm$^{-3}$], $-4.5 \le \log U \le 0$. Given the measured intensity ratios for xA quasars, {\em Cloudy} simulations  show convergence toward a well-defined value of $\log$ ($n_\mathrm{H}U$) \cite{negreteetal12,martinez-aldamaetal18}.  UV diagnostic ratios in the plane ionization parameter versus density indicate  extremely high $n_\mathrm{H} ~ 10^{12.5 - 13}$ cm$^{-3}$, extremely low $\log U \sim -2.5 - 3$ (Fig. \ref{fig:diag}). 
Note the orthogonal information  provided by the AlIII$\lambda$1860/SiIII]$\lambda$1892 that mainly depends on density. The left and right panel differ because of chemical abundances: the case with five times solar metallicity plus overabundance of Si and Al produces better agreement, displacing the solution toward lower density and higher ionization. Nonetheless, the product $U n_\mathrm{H}$ remains fairly constant.  Diagnostic ratios sensible to chemical composition suggest high metallicity. { The metallicities in the quasar BLR gas   are a function of the ST along the MS: relatively low (solar or slightly sub-solar in extreme Pop. B sources (as estimated recently for NGC 1275 \cite{punslyetal18a}), and  relatively high for typical Pop. A quasars with moderate FeII emission ($Z \sim 5 - 10Z_{\odot}$, \cite{shinetal13,sulenticetal14}). If the diagnostic ratios are interpreted in terms of scaled $Z_{\odot}$, they may reach  $Z \gtrsim 20 Z_{\odot}$, even $Z \sim 100 Z_{\odot}$ for xA quasars \cite{martinez-aldamaetal18}.  $Z$\  values as high as $Z \sim 100 Z_{\odot}$ are likely to be unphysical,} and suggest relative abundances of elements deviating from solar values, as assumed in the previous example. The analysis of the gas chemical composition in the BLR of xA source has just begun. However,  high or non-solar $Z$\ are in line with the idea of xA sources being high accretors surrounded by huge amount of gas and a circum-nuclear star forming system, possibly with a top-heavy initial mass function \cite{negreteetal12}. The high $n_\mathrm{H}$ is consistent with the low CIII]$\lambda$1909 emission that becomes undetectable in some cases. While in Pop. A and B we find evidence of ionization stratification within the low-ionization part of the BLR (\cite{petersonwandel00,petersonwandel99}, \cite{gaskell09b} and references therein), xA sources show intensity ratios that are consistent with a very dense ``remnant'' of the BLR, perhaps after lower density gas has been ablated away by radiation forces.

\section{xA quasars as Eddington standard candles }
\label{esc}

There are several key elements that make it possible to exploit xA quasars as Eddington standard candles. 

The first is the similarity of their spectra and hence of the physical condition in the mildly-ionized gas that is emitting the LILs. Line intensity  ratios are similar (they scatter around a constant average with small dispersion). Since the line emitting gas is photoionized, intensity line ratios depend strongly on the ionizing continuum SED. Thus, also the ionizing SED is constrained within a small scatter. We remark that this is not true for the general population of quasars that show differences in line equivalent width and intensity ratios larger than an order of magnitude along the MS. 

{  The mass reservoir in all xA sources is sufficient to ensure a very high accretion rate (possibly super-Eddington) that yields a  radiative output close to the Eddington limit.  The similarity of the SED and the presence of high rates  of circumnuclear and galactic  star formation as revealed by Spitzer \cite{sanietal10}, has led to the conjecture that xA sources may be in particular stage of a quasar development, as mentioned above.}

The second key element is the existence of a virialized low-ionization sub-region (possibly the accretion disk itself). This region coexists with outflowing gas even at extreme $L\gtrsim10^{48}$ erg s$^{-1}$ and highest Eddington ratios, but is kinematically distinguishable on the basis of inter-line shifts between LILs and HILs, for exampleH$\beta$\ and CIV$\lambda$1549.  

In addition,  xA quasars show extreme $L/L_\mathrm{Edd}$   along the MS with small dispersion.  If the Eddington ratio  is known, and constant, then $\lamed = L/L_\mathrm{Edd} \propto L/M_\mathrm{BH}$. Accretion disk theory teaches low radiative efficiency at high accretion rate, and that $\lamed$ saturates toward a limiting values (\cite{abramowiczetal88,mineshigeetal00,abramowiczstaub14} and references therein). Therefore,  empirical evidence (the xA class of sources easily identified by their self-similar properties,  scatters around a well-defined, extremal $\lamed$)  and theoretical support (the saturation of the radiative output per unit $M_\mathrm{BH}$)  justified the consideration of xA sources potential ESCs. 

\subsection{Virial luminosity}
\label{virial}

The use of xA sources as Eddington standard candles requires several steps which should considered carefully. 

\begin{enumerate}
\item The first step is the actual estimate of the accretion luminosity via a virial broadening estimator (VBE). The luminosity can be written as 

\begin{equation}
L \propto \lamed M_\mathrm{BH} \propto \lamed r_\mathrm{BLR} (\delta v)^{2}
\label{eq:vir0}
\end{equation}

assuming virial motions of the low-ionionization part of the broad-line region (BLR).  The $\delta v$\ stands for a suitable VBE, usually the width of a convenient LIL (in practice, the FWHM of H$\beta$\ or even Pa$\alpha$, \cite{lafrancaetal14}). 

\item The  $r_\mathrm{BLR}$\ can be estimated from the inversion of Eq. \ref{eq:ip} { \cite{negreteetal12,negreteetal13}}, taking again advantage that the ionizing photon flux shows a small scatter around a well defined value. In addition, another key assumption is that 

\begin{equation}
r_\mathrm{BLR} \propto \left(\frac{L}{n_\mathrm{H}U}\right)^{\frac{1}{2}}.
\label{eq:r}
\end{equation}
 
 Eq. \ref{eq:r} implies that $r_\mathrm{BLR}$\ scales with the square root of the luminosity. This is needed to preserve the $U$\ parameter. If $U$ were going to change then the spectrum would also change as a function of luminosity. This is not evident comparing spectra over a wide luminosity range (4.5 dex), although some second order effects are possible.\footnote{The maximum temperature of the accretion disk is $\propto M_\mathrm{BH}^{-\frac{1}{2}}$; the SED is expected to become softer at high $M_\mathrm{BH}$, but this effect has not been detected yet at a high confidence level.} 

\item We can therefore write the virial luminosity as 

\begin{equation}
L \propto \lamed \left(\frac{L}{n_\mathrm{H}U}\right)^{\frac{1}{2}} (\delta v)^{2}.
\end{equation}

Making explicit the dependence of the number of ionizing photons on the SED, the virial luminosity becomes:  

\begin{equation}
L \approx 7.8 \cdot 10^{44} \frac{\lamed_{1}^{2} \kappa_{0.5} f_{2}^{2}}{\tilde{\nu}_{2.42 \cdot 10^{16}}}  \frac{1}{\left({n_\mathrm{H}U}\right)_{9.6}} (\delta v_{1000})^{4} 
\label{eq:vir}
\end{equation}

where $\kappa$\ is the fraction of ionizing luminosity scaled to 0.5,   $\tilde{\nu}$
the average frequency of ionizing photons scaled to $2.42 \cdot 10^{16}$ Hz, and $\left({n_\mathrm{H}U}\right)$ to $10^{9.6}$. \end{enumerate}

Eq. \ref{eq:vir} is analogous to the Tully-Fisher \cite{tullyfisher77} and the early formulation of the Faber-Jackson \cite{faberjackson76} laws for galaxies. Eq. \ref{eq:vir} is applicable to xA quasars with Eddington ratio $\lamed \sim 1$ and  dispersion $\delta \lamed  \ll 1$,  but in principle could be used to for every sample of quasars whose $\lamed$\ is in a very restricted range.  


\section{Selection of Eddington standard candles}
\label{selec}
                                                                                                                                                                                                                         
Selection criteria are based on emission line intensity ratios which are extreme along the quasar MS: 
\begin{enumerate}
\item $R_\mathrm{FeII}$ \ $> 1.0$;
\item UV AlIII$\lambda$1860/ SiIII]$\lambda$1892$>0.5$;
\item SiIII]$\lambda$1892/ CIII]$\lambda$1909$>1$
\end{enumerate}

{ \cite{marzianisulentic14}. The first criterion can be easily applied to optical spectra of a large survey such as the SDSS for sources at $z \lesssim 1$. The second and third criterion can be applied to sources at $1 \lesssim z \lesssim 4.5$\ for which the 1900 blend lines are shifted into the optical and near IR domains. }  UV and optical selection criterions are believed to be equivalent. Due to a small sample size at low $z$\ for which rest-frame optical and UV spectra are available, further testing is needed. 

\section{Tentative applications to cosmology and the future perspectives}
\label{cosmo}

\begin{figure}
\begin{minipage}[t]{0.4\linewidth}
\centering
\includegraphics[scale=0.175]{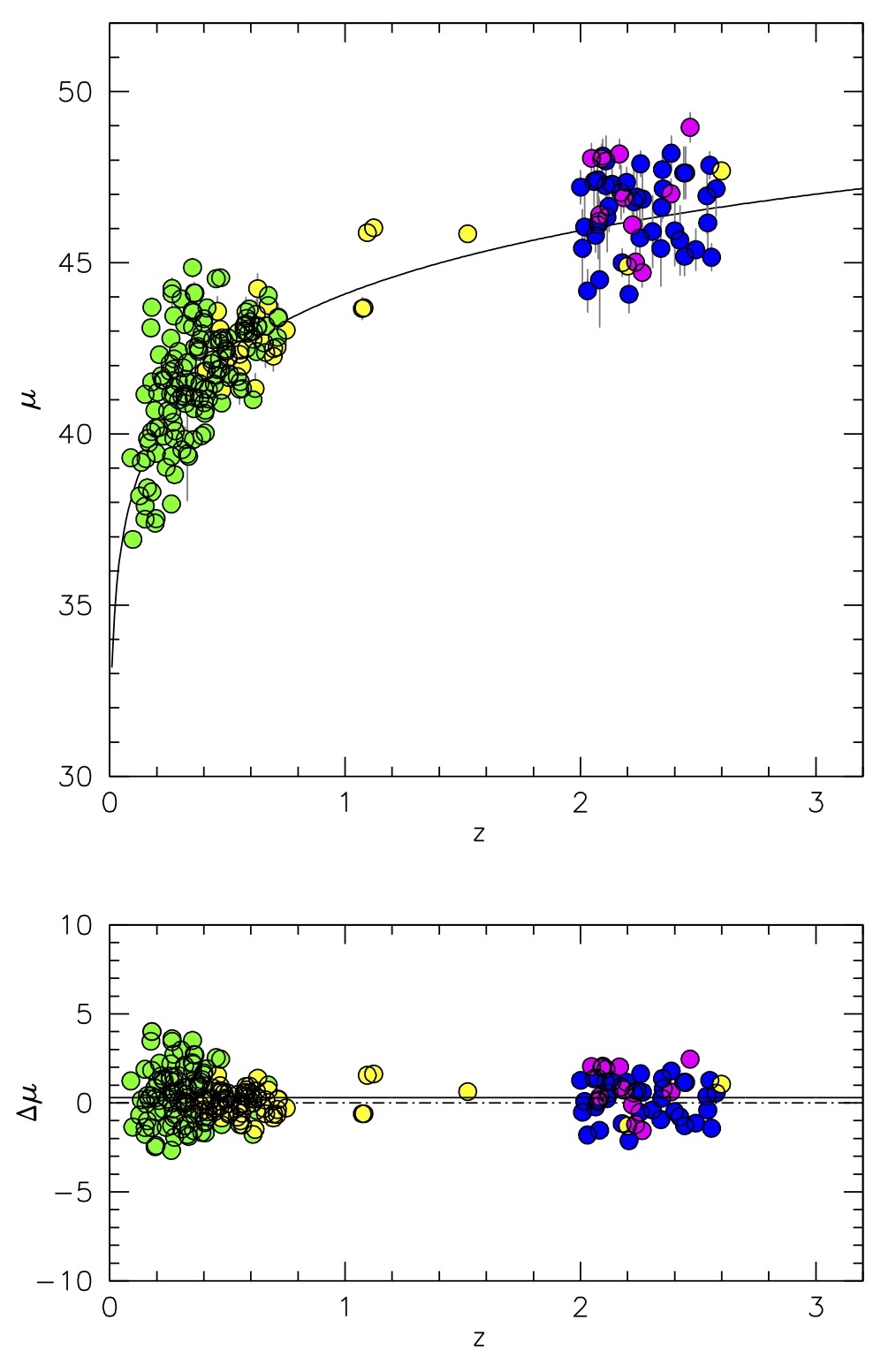}
\end{minipage}
\hspace{0.25cm}
\begin{minipage}[thp!]{0.55\linewidth}
\centering\vspace{-11cm}
\caption{Hubble diagram distance modulus $\mu$ vs $z$\ obtained from the analysis of the \cite{marzianisulentic14} data (yellow: H$\beta$, navy blue: Al{\sc iii}$\lambda$1860 and Si{\sc iii}]$\lambda$1892) supplemented by new H$\beta$\ measurements from the SDSS obtained in this work (green) and from GTC observations of \citet{martinez-aldamaetal17} (magenta).   The lower panel shows the distance modulus residuals with respect to concordance cosmology. The filled line in the upper panel is the $\mu$($z$) expected from $\Lambda$CDM cosmology. The filled line in the lower panel represents a lsq fit to the residuals as a function of $z$. The Figure is an updated version of Fig. 1 of \citet{marzianietal17}.  \label{fig:hub} }
\end{minipage}
\end{figure}

Preliminary results were collected from 3  quasar samples (62 sources in total), unevenly covering the redshift range $0.4 \lesssim z \lesssim 2.6$. For redshift $z \gtrsim 2.$ the UV AlIII$\lambda$1860 FWHM was used as a VBE for the rest-frame UV range, save a few cases for which H$\beta$ was availabele. This explorative application to cosmology yielded results consistent with concordance cosmology, and allowed the exclusion of some extreme cosmologies \cite{marzianisulentic14}. A more recent application involved the \cite{marzianisulentic14} sample, along with the H$\beta$\ sample of \cite{negreteetal18} and preliminary measurements from \cite{martinez-aldamaetal18}. The resulting Hubble diagram  is shown in Fig. \ref{fig:hub}. The plots in Fig. \ref{fig:hub} involve $\approx 220$  sources and indicate a scatter $\delta \mu \approx $ 1.2 mag. The  slope of the residuals ($b \approx -0.002 \pm  0.104$)\ is not significantly different from 0, indicating good statistical agreement between luminosities derived from concordance cosmological parameters and from the virial equation. The Hubble diagram of  Fig. \ref{fig:hub}  confirms the conceptual validity of the virial luminosity relation, Eq. \ref{eq:vir}.

\begin{figure}[H]
\begin{minipage}[t]{0.4\linewidth}
\centering
\includegraphics[width=7 cm]{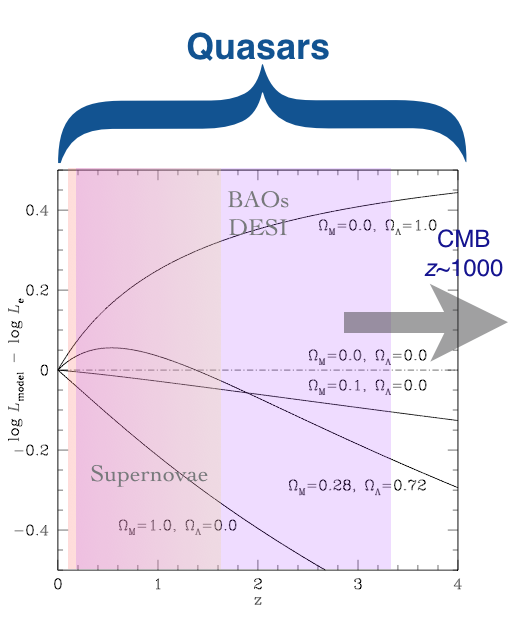}
\end{minipage}\hspace{1cm}
\begin{minipage}[t]{0.5\linewidth}
\centering
\vspace{-7cm}
\caption{Luminosity difference with respect to an empty Universe for several cosmological models, identified by their values of $\Omega_{M}$ and $\Omega_{\Lambda}$. The domain of supernov\ae\ and of the baryonic acoustic oscillations (within the expectation of the future DESI survey, \cite{levietal13}) are shown. \label{fig:scheme}}
\end{minipage}
\end{figure}

Mock samples of several hundreds of objects, even with significant dispersion in luminosity with rms($\log L$) = 0.2 -- 0.3, indicate that quasars covering the redshift range between 0 and 3 (i.e., a range of cosmic epochs from now to 2 Gyr since the Big Bang) could yield significant constraints on the cosmological parameters. { A synthetic sample of 200 sources uniformly distributed in the redshift range 0 -- 3 with a scatter of 0.2 dex yields  $\Omega_\mathrm{M} \approx 0.28 \pm 0.02$\ at 1$\sigma$\ confidence level, assuming $H_{0} = 70$ km s$^{-1}$\ Mpc$^{-1}$, and flatness ($\Omega_{M}+\Omega_{\Lambda}$=1). If $\Omega_{M}+\Omega_{\Lambda}$\ is unconstrained, $\Omega_\mathrm{M} \approx 0.30^{+0.12}_{-0.09}$\ at 1$\sigma$ confidence level \cite{marzianisulentic14}. } 
The comparison between the constraints set  by supernova  surveys  and by a mock sample of 400 quasars with rms = 0.3 dex in $\log L$\ shows the potential ability of the quasar sample to better constrain $\Omega_{M}$\ \cite{marzianisulentic14a}. The scheme of Fig. \ref{fig:scheme} illustrates the difference in sensitivity to cosmological parameters over the redshift range 0 -- 4: supernov\ae\ are sensitive to $\Omega_{\Lambda}$\ since the effect of $\Omega_{\Lambda}$, in a concordance cosmology scenario, became appreciable only at relatively recent cosmic epochs. High redshift quasars provide information on a redshift range where the expansion of the Universe was still being decelerated by the effect of $\Omega_{M}$, a range that is not yet covered by any standard ruler or candle.  

\subsection{Error budget}

The large scatter in the luminosity estimates is apparently daunting in the epoch of precision cosmology.  Statistical errors can be  reduced to rms $\approx 0.2$ dex in $L$\ by increasing numbers, collecting large samples ($\sim$ 500 quasars).  

In the case of xA quasars, the SED cannot vary much since spectra of xA quasars are almost identical in terms of line ratios (a second order effect \cite{shemmerlieber15}  not  yet detected as significant in the data considered by \cite{marzianisulentic14,martinez-aldamaetal18}  may become significant with larger samples).   The scaling $r_\mathrm{BLR} \propto L^{0.5}$\ should hold  strictly: a small deviation would imply a systematic change in the ionization parameter and hence of ST with luminosity.

A simplified error budget for statistical errors \cite{marzianisulentic14} clearly indicates that the virial luminosity estimate is affected by the VBE uncertainties (i.e., FWHM measurement uncertainties) which enter with the fourth power in Eq. \ref{eq:vir}. Orientation effects are expected to be determinant in the FWHM uncertainties, as they can contribute 0.3 dex if H$\beta$ or any other line used as a VBE is emitted in a highly-flattened configuration.\footnote{{ The $M_\mathrm{BH}$\ is not computed explicitly for the estimate of  $L$ following Eq. \ref{eq:vir}. However,  the VBE uncertainty associated with orientation is the main source of uncertainty on $M_\mathrm{BH}$\ for the xA sample. This is most likely the case also for the general quasar population \cite{jarvismclure06,marzianietal17d}.}}  { Modeling the effect of orientation by computing the difference between the $L$  from concordance cosmology and the virial luminosity indeed reduces the sample standard deviation by a factor $\approx$ 5 to $\approx$ 0.2 mag, and accounts for most of the rms  $\approx$ 0.4 dex in the virial luminosity estimates of the sample shown in Fig. \ref{fig:hub} \cite{negreteetal18}.} The rms $\approx$ 0.2 mag value is comparable to the uncertainty in supernova magnitude measurements. Work is in progress in order to make viewing angle estimates of { xA quasars} usable for cosmology.


\section{Conclusions}

{ This paper provided an overview of the physical  conditions in the broad line emitting region of extreme spectral types of type-1 quasars (the extreme Pop. A). There is strong evidence that xA sources are radiating close to their Eddington limit (i.e., with Eddington ratio scattering around a well-defined value), at high accretion rates. Their physical properties appear to be very stable across a very wide range of luminosity, 4 -- 5 dex. 
The assumption of a constant  $\lamed$\  makes it possible to write a relation between luminosity and virial broadening, analogous to the one expressed by the Tully-Fisher and the early formulation of Faber-Jackson laws.}  

The scatter in the Hubble diagram obtained from  virial luminosity estimates is still very high, about 1 mag (although comparable to the scatter from a method based on the non-linear relation between the X-ray and the UV emission of quasars \cite{risalitilusso15}). Very large samples are needed for reduction of scatter (and statistical error). In addition, the inter-calibration of rest-frame visual and UV properties and their dependence on $L$\ (expect systematic errors!) needs to be extended by dedicated observations of xA sources covering the rest frame UV and visual range. Simulations of statistical and systematic effects which influence the estimates of the cosmic parameters are also needed.

In principle, Eddington standard candles can cover a range of distances where the metric of the Universe has not been ``charted'' as yet to retrieve an independent estimate of $\Omega_\mathrm{M}$. If samples with uniform coverage over a wide range of redshift would become available, xA sources could  also address the physics of accelerated expansion (i.e., provide measurements of the dark energy equation of state).




\vspace{6pt} 



\authorcontributions{All authors significantly contributed  to the papers on which this review is based.}

\funding{PM  wishes to thank the SOC of the SPIG 2018 meeting for inviting the topical lecture on which this paper is based, and acknowledges the Programa de Estancias de Investigaci\'on (PREI) No. DGAP/DFA/2192/2018 of UNAM, where this paper was written.  The relevant research is part of the project 176001 ''Astrophysical spectroscopy of extragalactic objects``  and 176003 ``Gravitation and the large scale structure of the Universe" supported by the Ministry of Education, Science and Technological Development of the Republic of Serbia. MLMA acknowledges a CONACyT postdoctoral fellowship. AdO, MLMA  acknowledge financial support from the Spanish Ministry for Economy and Competitiveness through grants AYA2013-42227-P and AYA2016-76682-C3-1-P. MLMA, PM and MDO acknowledge funding from  the INAF PRIN-SKA 2017 program 1.05.01.88.04. DD and AN acknowledge support from CONACyT through  grant CB221398. DD and AN thank also for support from grant IN108716
53 PAPIIT, UNAM. }

\conflictsofinterest{The authors declare no conflict of interest.} 

\abbreviations{The following abbreviations are used in this manuscript:\\

\noindent 
\begin{tabular}{@{}ll}
AGN& active galactic nucleus\\
BLR& Broad Line Region\\
DESI& Dark Energy Spectroscopic Instrument \\
ESC& Eddington standard candles\\
HIL& High-ionization line\\
LIL& Low-ionization line\\
MDPI& Multidisciplinary Digital Publishing Institute\\
MS& Main Sequence\\
NLSy1 & Narrow-Line Seyfert 1\\
SDSS & Sloan Digital Sly Survey\\
VBE& Virial Broadening Estimator\\
\end{tabular}}


\reftitle{References}
\end{document}